\documentclass[a4paper,12pt]{article}
\usepackage{cite}
\usepackage{amssymb,amsmath}
\newcommand{\be}{\begin{equation}}
\newcommand{\ee}{\end{equation}}
\newcommand{\bea}{\begin{eqnarray}}

\newcommand{\eea}{\end{eqnarray}}

\begin{document}

\title{Yet More on the ``Universal'' Quantum Area  
Spectrum~\footnote{Alternative Title: {\it Comment on 
``A Note on the Lower Bound of Black Hole Area Change in Tunneling
  Formalism''}}}

\author{A.J.M. Medved \\ \\
Physics Department \\
University of Seoul \\
Seoul 130-743\\
Korea \\
E-Mail(1): allan@physics.uos.ac.kr \\
E-Mail(2): joey\_medved@yahoo.com \\ \\}

\maketitle
\begin{abstract}

We briefly comment on the quantum area spectra of black holes, paying particular
attention to the size of the spacing between adjacent spectral levels. It has 
previously been conjectured that this spacing is uniform with a universal value 
of $8\pi$ in Planck units. In spite of a recent claim to the contrary 
\cite{BMV}, we argue that this particular value remains, by far, the most 
qualified candidate for a universal area gap.

\end{abstract}
\newpage
\section*{ }

Ever since Bekenstein \cite{bek-1} proposed that a black hole should have a quantum spectrum for its horizon area,
there has been a significant amount of debate on the topic. The discourse has covered both the form of the spectra --- evenly spaced or otherwise ({\it e.g.}, \cite{rovelli}) ---
and, if uniformly spaced as per Bekenstein's original account, then
the size of the gap between adjacent levels ({\it e.g.}, \cite{hod-1}). Let us, for the sake of current considerations,
take the evenly spaced form as a given and focus our attention on the second issue.

Formally speaking, we are then asking as to the size of the
dimensionless parameter $\gamma$ when the spectrum for the area operator
($\;A_n \equiv <{\hat A}> \;$) is expressed as
\be
A_{n}\;=\; A_0 \;+\; \gamma l_P^2 n\;\;\;{\rm where}\;\;\;n=0,1,2,\ldots \; .
\label{1}
\ee
Here, $l_P$ is the Planck length~\footnote{We have committed to four
dimensions, with the usual disclaimer that generalizations are readily
attainable. Also note that, for future reference, the speed of light and
the Boltzmann
constant are always  set to unity.} and
$A_0$  allows for the possibility of a non-vanishing zero-point term.
Given that $l_P$ is the length scale at which quantum gravity sets in,
it is naturally expected that $\gamma$ is of the order of unity. However,
without further inputs, it is difficult to be more concise on its value.

If one wishes to pursue the matter, the two viable avenues (that the current author
is aware of) are
to either appeal to a specific theory of quantum gravity or
to follow some semi-classical line of reasoning. An interesting example of
the second option is seen in a very recently submitted
letter by Banerjee {\it et. al.} \cite{BMV}.
Closely following Bekenstein's  legendary treatment of
 a black hole  absorbing a quantum particle in the vicinity of its horizon   \cite{bek-4},
these authors arrive at the following bound (although  translated into our 
conventions):
\be
\gamma \; \geq \frac{8\pi}{\hbar} \epsilon \Delta \hat{X} <\hat{E}> \;.
\label{2}
\ee
Here, $\hat{E}$ and $\hat{X}$ are quantum operators for the particle's energy and displacement from the  horizon,
whereas $\epsilon$
is a dimensionless ``fudge factor'' reflecting the rather ambiguous nature of assimilation
by a black hole. (Meanwhile, $\Delta$ and $<~>$ always maintain 
their usual meanings of
quantum uncertainty and expectation value.)

The cited authors then appeal to a variation \cite{BM,BM2}  of the  often-discussed tunneling mechanism \cite{Tun}
(although knowledge of Hawking's thermal spectrum for  black hole radiation \cite{haw-1} and
 a text book on statistical mechanics would have served just as well)
to deduce that
\be
<\hat{E}>\; =\; T_H \;,
\label{3}
\ee
with $T_H$ denoting the Hawking temperature of the black hole.
Additionally, for later use,
 $\;<\hat{E^2}>=2T_H^2\;$, so that
\be
\Delta \hat{E} \;=\; \sqrt{<\hat{E^2}>-<\hat{E}>^2} \;=\; T_H\;.
\label{4}
\ee
Substituting Eq.~(\ref{3}) into (\ref{2}), they then obtain
\be
\gamma \; \geq \frac {8\pi}{\hbar} \epsilon \Delta \hat{X} T_H \;.
\label{5}
\ee
So far, so good.

The critical steps are now upon us. The stated authors use the quantum
uncertainty principle $\;\Delta \hat{X} \Delta \hat{P}\geq \hbar\;$ (with 
$P$ representing the particle's momentum operator) and the obvious bound $\;\Delta \hat{E}\geq \Delta \hat{P}\;$
(or taken as an equality for a massless particle) to argue that
\be
\Delta \hat{X}\; \geq \; \frac{\hbar}{\Delta \hat{E}}\;
\label{6}
\ee
and then, by virtue of  Eq.~(\ref{4}),
\be
\Delta \hat{X}\; \geq \; \frac{\hbar}{T_H}\;.
\label{7}
\ee
Substituting this result into Eq.~(\ref{5}), they end up with
\be
\gamma \; \geq 8\pi \epsilon \;.
\label{8}
\ee

Had the authors  stopped here, it would be a  perfectly reasonable
deduction. And, given the
ambiguity that is inherent through $\epsilon$, almost
certainly a true observation (not to mention reminiscent 
of  Bekenstein's \cite{bek-4}). 
However, they then proceed to
 {\it double down}~\footnote{For any confused readers,
  this terminology is  borrowed from the casino game ``Black Jack''. Doubling down
  is a strategic option that enables one to double his or her potential winnings
  at the cost of doubling the potentiality for loss.} on their analysis
by attempting to fix $\epsilon$ via the first law of black hole
mechanics \cite{boys}.  More to the point,  this law can be expressed
(in its simplest form)  as
\be
T_H \frac {\delta A}{4 l_P^2}\;=\; \delta M \;,
\label{9}
\ee
where $M$ is the mass or  rest energy of the black hole.
Let us rewrite this by
attributing $\delta M$ to the absorption of
the previously discussed particle and then suitably quantizing:
\be
T_H \frac {\gamma}{4} \;=\; <{\hat E}> \;.
\label{10}
\ee
Substituting Eq.~(\ref{3}) into the above expression
--- which is reasonable facsimile of that used by the 
discussed authors~\footnote{Actually,
the authors of \cite{BMV} take the change in black hole mass to be $\Delta {\hat E}$, which
does not seem quite right. However, since $\;\Delta {\hat E}= <{\hat E}>=T_H\;$, their
choice amounts to the same thing.} --- we 
finally arrive at the authors claim of
\be
\gamma \; = \;4 \;.
\label{11}
\ee
Or, in other words, $\;\epsilon= 1/2\pi\;$  
with Eq.~(\ref{8}) now taken to be a  strict  equality.

 Although this all seems reasonable enough,
 the logic is unfortunately flawed, as we now explain: The identification
 $\; \delta M = <\hat{E}>=T_H \;$
 might be true if one were endeavoring to calculate the thermal
 or {\it canonical} fluctuations in the horizon area.
  Indeed, the very notion of evaluating variations in energy at a
 fixed value of temperature (in this case $T=T_H$) only
 makes sense in a canonical setting. However, this is not what is
 meant (at least not purposefully) when one talks about the quantum area spectrum of a black hole.
 Eq.~(\ref{1}) is, rather,  meant to be the {\it fundamental} quantum spectrum
 for a black hole, which implies that the appropriate setting
 is, in actuality, a {\it microcanonical} 
one.~\footnote{Although long advocated by
 the current author ({\it e.g.}, \cite{med-y}), 
G. Gour was the first to emphasize the importance
 of distinguishing between the canonical and microcanonical contributions
 to the area spectrum \cite{gilad}.}
  That is,
 one should fix the energy of the system and then inquire as to how
 the spectral levels of a given quantity (in this case the area) are
 distributed. In all practicality, thermal fluctuations would be
 present and, likely, blur the original spectral lines. But such
 fluctuations are not relevant to questions about the   fundamental
 nature of the quantized geometry.

 Then, is it possible to be more definitive about the spacing parameter
 $\gamma$? To this end,
 let us start with the simplest case of a Schwarzschild black hole, for
 which there is only one relevant length scale: the horizon radius or,
 equivalently, the  inverse of the Hawking temperature. Hence, it
 can be expected on dimensional grounds  that $\;\delta M \sim T_H\;$,
 so that $\gamma$ is of order unity. Now, insofar as  
 $\gamma$ truly has the status of a  universal parameter 
(in accordance with Occam's razor
if nothing else),
 it would follow that the very same order-unity value carries through to more
 elaborate scenarios; including   black holes of the spinning, charged 
and/or ``hairy'' 
variety.~\footnote{Even more elaborate is  when the gravitational
theory differs from Einstein's. In this case, $\gamma$ should
be regarded as the spacing between the spectral levels of the operator
$\;{\hat S_W}/4l_P^2\;$, where $S_W$ is meant to represent Wald's geometric
or Noether-charge  entropy \cite{wald}. Given a generic theory of gravity, 
the spectrum  for this quantity is, 
as made clear in 
\cite{pad-2,med-rop}, the unequivocal
  analogue to the area spectrum.}

 We can not, however, be more specific than this
 without further inputs, which generally (if not inevitably) necessitates
 further assumptions about what constitutes a viable quantum theory of gravity.
 Nevertheless, many treatments have independently produced the same value
 of  $\;\gamma = 8\pi\;$ --- see \cite{med} and references therein. More recent examples include a quantization
procedure proposed by Ropotenko \cite{rop}, a refinement thereof \cite{med-rop} and,
perhaps most persuasively, Maggiore's reinterpretation 
\cite{mag}~\footnote{Also see, for instance,\cite{x1,x2,x3}.}  
of the renowned Hod conjecture \cite{hod-1}.
 In addition, it is probably  worth mentioning
  the ``emergent gravity'' conjectures of
  Padmanabhan \cite{pad} and
  Verlinde \cite{ver}. In this context, one finds that the unique choice
 of $\;\Delta S =2\pi\;$ (where $\Delta S$ is the minimal change in the entropy that is responsible
 for gravity) leads almost 
miraculously~\footnote{Or, perhaps, coincidentally.}  to 
Newton's second law of mechanics and Newton's law of gravitation, amongst others.
Translated in terms of the black hole area--entropy law \cite{bek-x,haw-1}, this becomes $\;\Delta A= 8\pi l_P^2\;$
or, once again, $\gamma =8\pi.$

It may be  true that all of the studies  returning $\gamma=8\pi$ are limited contextually by their scope
and/or technically through their assumptions (both explicit and implied).
So it is still feasible that all of these studies are simply wrong and $\gamma$ is not
$8\pi$ after all. For that matter, $\gamma$ may not even be universal.
Nevertheless,
taken as a whole, the overall body of evidence is pretty compelling.

A  possible counter-example, as pointed out by Banerjee {\it et. al.} \cite{BMV}, could be Hod's generalization \cite{hod-xx} of Bekenstein's calculation \cite{bek-4}. After revising the analysis to that of  a charged particle being absorbed by a charged (Reissner--Nordstrom) black hole, Hod  advocated  for the contrary result 
of  $\;\gamma=4\;$. However,
what Hod actually formulated  was a lower bound, so the more accurate statement is $\;\gamma \geq 4\;$, which
is in no way contradictory.

In conclusion, we assert that (I) $\;\gamma=8\pi\;$ is still, by far, 
the most qualified
candidate for a universal area spacing (if any) and (II) Banerjee {\it et. al.} have provided no evidence
in \cite{BMV}  that casts dispersion
upon the first claim.

\section*{Acknowledgments}
The author's research is financially supported by the University of
Seoul.

\end{document}